\documentclass[11pt]{article}
\usepackage{amsmath,amsbsy,amsfonts,amssymb,graphics,epsfig,float,mathrsfs,amsthm,braket}
\usepackage{ifsym}
\usepackage{psfrag}
\usepackage{color}
\usepackage[normalem]{ulem}

\newcommand{\hankel}[1]{\tilde{#1}}

\newcommand{\Es}{E_{\mathrm{subs}}}
\newcommand{\Estar}{E_{\mathrm{subs}}^\ast}
\newcommand{\Ef}{E_{\mathrm{film}}}
\newcommand{\Ustrain}{U_{\mathrm{subs}}}
\newcommand{\Uelast}{U_{\mathrm{elast}}}
\newcommand{\Uindent}{U_{\mathrm{ind}}}

\newcommand{\rind}{r_{\mathrm{ind}}}
\newcommand{\Rind}{\rho_{\mathrm{ind}}}

\newcommand{\ts}{t_{\mathrm{subs}}}
\newcommand{\tf}{t_{\mathrm{film}}}
\newcommand{\lstar}{\ell_\ast}

\newcommand{\tkappa}{\hat{\kappa}}

\newcommand{\upd}{\mathrm{d}}
\newcommand{\beq}{\begin{equation}}
\newcommand{\eeq}{\end{equation}}

\newcommand{\s}[1]{{\textsf{\textbf{#1}}}}

\begin{document}
\title{\s{Cloaking by coating: How effectively does a thin, stiff coating hide a soft substrate?}}
\author{ \textsf{Finn Box$^\dagger$, Cyprien Jacquemot$^\dagger$, Mokhtar Adda-Bedia$^\ddagger$ and Dominic Vella$^\dagger$}\\ 
{\it$^\dagger$Mathematical Institute, University of Oxford, Woodstock Rd, Oxford, OX2 6GG, UK}\\
{\it$^\ddagger$Universit\'{e} de Lyon, Ecole Normale Sup\'{e}rieure de Lyon, Universit\'{e} Claude Bernard,}\\
{\it CNRS, Laboratoire de Physique, F-69342 Lyon, France}}

\maketitle
\hrule\vskip 6pt
\begin{abstract}
From human tissue to fruits, many soft materials are coated by a thin layer of a stiffer material.  While the primary role of such a coating is often to protect the softer material, the thin, stiff coating also has an important effect on the mechanical behaviour of the composite material, making it appear significantly stiffer than the underlying material.   We study this cloaking effect of a coating for the particular case of indentation tests, which measure the `firmness' of the composite solid: we use a combination of theory and experiment to characterize the firmness quantitatively. We find that the indenter size plays a key role in determining the effectiveness of cloaking: small indenters feel a mixture of the material properties of the coating and of the substrate, while large indenters sense  largely the unadulterated substrate. 
\end{abstract}
\vskip 6pt
\hrule

\maketitle

\section{Introduction}

How does one tell when a piece of fruit is ripe? While for fruits such as tomatoes and bananas colour alone is a reliable indicator of ripeness \cite{Mizrach1992,Scimeca2019},  everyday experience suggests that for fruits including plums \cite{Plocharski2003} and mangoes \cite{Polderdijk2000,Scimeca2019}, one must instead  `poke' the fruit: if the fruit is soft then it is ripe, while if  relatively stiff the flesh is not yet ripe. Of course, how soft is soft enough depends on the type of fruit and is knowledge gained by experience. As well as being of importance to consumers assessing the ripeness of fruit in shops and at home, measurements of fruit ripeness is also important to producers\cite{Abbott1999}. A common strategy producers use for measurements of  ripeness is a mechanized version of the poking test used by consumers: the force required to impose a given indentation depth via a cylindrical punch is measured and the resulting stiffness is then correlated to the ripeness. Particular protocols have been proposed for fruits including apples \cite{Duprat1995,Grotte2001}, plums \cite{Plocharski2003}, pumpkins \cite{Emadi2005}, mangoes \cite{Polderdijk2000,Scimeca2019},  oranges and tomatoes \cite{Mizrach1992}.

A feature common to both industrial and domestic tests of fruit ripeness is that fruits are  usually protected by a thin, but stiff, skin protecting the softer flesh \cite{Wang2017}. The industrial literature generally recommends peeling fruit first to avoid anomalously large stiffness measurements \cite{Grotte2001, Emadi2005} --- thereby sacrificing one fruit as a representative of a large batch. While this sacrifice may work in an industrial setting, it is not practical for the consumer who needs a non-destructive test.  The question then is: how is the measured  stiffness affected by the large  stiffness of the thin skin? To what extent is the stiffness of the flesh (the quantity of interest) cloaked by the stiffness of the skin?

Similar scenarios arise in many problems in soft matter: stiff, thin layers (including graphene) are adhered to thicker soft substrates in applications including membrane separation \cite{Paraense2017}, photovoltaics \cite{Kim2012} and flexible electronics \cite{Kaltenbrunner2013}. In such applications, as in the case of many fruits, the Young's modulus of the coating, $\Ef$, is significantly larger than that of the substrate, $\Es$, i.e.~$\Ef/\Es\gg1$, but the ratio of their thicknesses $\tf/\ts\ll1$ ---- how does the composite material behave? In this paper, we  seek to understand how these composite materials respond to indentation, focussing on understanding the composite stiffness that is familiar from the preceding discussion of poking fruit.

The deformation of an uncoated  elastic half-space caused by a normal pressure distribution in some region, but otherwise unloaded is an old problem in mechanics. This problem was first considered by Boussinesq \cite{Boussinesq1878} and  two variants of it are now referred to as the `Boussinesq problem'\cite{Love1939}: in the first variant, a known pressure distribution is applied over a small region and the induced vertical deformation calculated. In the second variant, a known normal displacement is imposed in some region but the normal pressure distribution within that region must be determined. This second variant of Boussinesq's problem results in mixed boundary value problems \cite{Duffy2008}, which are, in general,  difficult to solve analytically and show features, such as stress singularities at the edge of the contact region\cite{Harding1945,Sneddon1995}, that are not present in the first variant of Boussinesq's problem. Nevertheless, Harding \& Sneddon \cite{Harding1945} provided solutions for the indentation of an uncoated substrate by indenters of particular profile; these were subsequently generalized to arbitrary axisymmetric indenter shapes by Sneddon \cite{Sneddon1965}. For a cylindrical indenter of radius $\rind$, these results suggest that the ratio of applied load and deflection is constant, corresponding to an indentation stiffness $\kappa\sim \Es\rind$ --- this result that will become a useful benchmark in this study, and is referred to as Hertz contact \cite{Johnson1987}.

The methods of Harding \& Sneddon \cite{Harding1945} and Sneddon \cite{Sneddon1965} cannot, however,  be generalized to the  coated substrate problem of interest here.  An alternative approach has therefore been to return to the first variant of Boussinesq's problem (assume a known spatial pressure distribution is applied and calculate the resulting deformation) \cite{Li1997,Liu2019}. The results of such calculations may be analytical (or require significantly simpler numerical calculations) but they come with the caveat that they do not truly represent the effect of a rigid indenter applied to the composite material.

 Some analytical progress for the coated problem has been made by considering particular asymptotic limits. For example, Yu \emph{et al.}~\cite{Yu1990} were motivated by indentation tests of thin films of ceramic-metal composites deposited on hard surfaces; their focus was therefore on understanding how the substrate properties  should be controlled to ensure that they do not unduly affect measurement of the thin film's properties by indentation. In such scenarios, the ratio of the moduli of the two layers is close to unity and so Gao \emph{et al.} \cite{Gao1992} developed asymptotic  results for the effective modulus of the composite exploiting the closeness in the ratio of the layers' moduli. These analytical results were then shown to be in good agreement with the numerical solutions provided that the shear moduli were within a factor of two of each other.

While the analytical approach of Gao \emph{et al.} \cite{Gao1992} is useful when the materials are similar in elastic modulus, many recent applications have significantly larger stiffness ratios. For example, a glassy layer might typically have $\Ef=O(1\mathrm{~GPa})$, while a soft polydimethylsiloxane (PDMS) substrate has \cite{Johnston2014} $\Es=O(1\mathrm{~MPa})$ (or even $\Es=O(10\mathrm{~kPa})$). In this case,  $\Ef/\Es\gtrsim10^3$ and so analytical approaches such as those of ref.~\cite{Gao1992} are no longer appropriate. More recent work has therefore focussed on providing numerical results for the effective modulus of the combined system with larger elastic mismatches $\Ef/\Es\neq O(1)$. For example, Perriot \& Barthel \cite{Perriot2004} provided numerical results for $10^{-2}\leq\Ef/\Es\leq 10^2$.

\begin{figure}
\centering
\includegraphics[width=0.65\columnwidth]{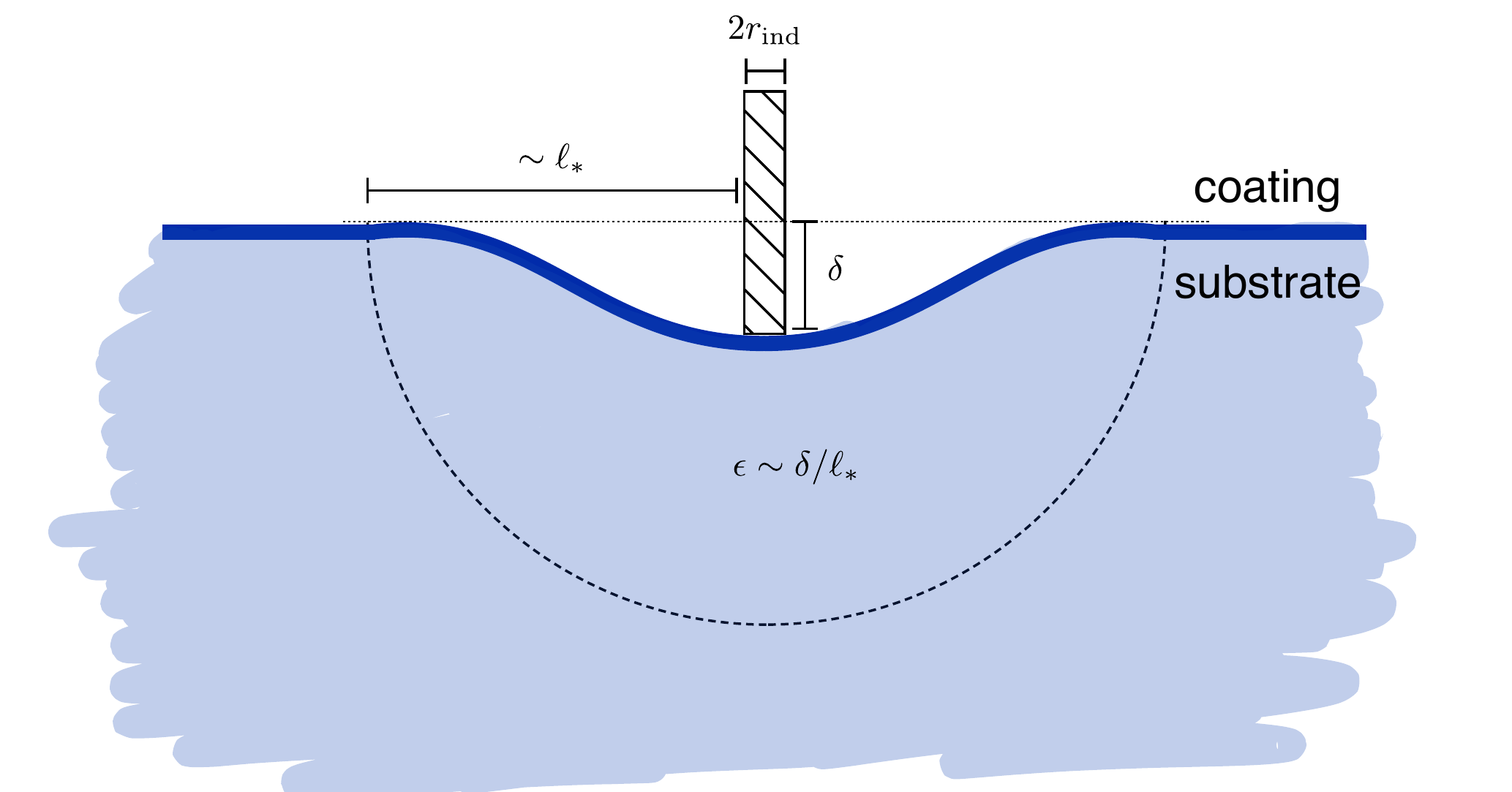}
\caption{A thin, stiff coating of a soft substrate is deformed by the application of a cylindrical indenter of radius $\rind$. The effect of the indentation is felt within the substrate through the imposed strain $\varepsilon\sim \delta/\lstar$, which penetrates a typical distance $\lstar$ throughout the substrate. In this paper we seek to determine the relationship between the applied indentation force, $F$, and the indentation depth $\delta$. This involves determining the characteristic lateral length scale $\lstar$ over which the substrate and coating are deformed by indentation.}
\label{fig:setup}
\end{figure}

In this paper, we present a model of the indentation of a coated soft substrate in which the effect of the coating is modelled as an elastic plate\cite{Ventsel2001} of bending stiffness $B=\Ef^\ast\tf^3/12$, with $\Ef^\ast=\Ef/(1-\nu_f^2)$ and $\nu_f$ the coating's Poisson ratio.  This approximation allows for some analytical progress to be made, as well as for some simplification of the problem to be solved numerically in situations where analytical progress is not possible. Moreover, this approximation is expected to be valid provided that the lateral length scale over which the coating is deformed, which we denote $\lstar$ as in fig.~\ref{fig:setup}, is very large compared to the thickness i.e.~$\lstar\gg\tf$.  However, the length scale $\lstar$ is not known \emph{a priori} and must be determined as part of the solution of the problem. We therefore turn to first understand the length scale $\lstar$ via a scaling analysis in \S\ref{sec:scalings}, before presenting experimental  (\S\ref{sec:ModelExpt}) and model  (\S\ref{sec:MathModel}) results.  We shall also compare  our results with those of previous works (summarized in table \ref{table:1}) in \S\ref{sec:Previous} before discussing the relevance of our results for the indentation of, among other things, fruits in \S\ref{sec:Discuss}. 

\begin{table*}[h]
\tiny
\centering
  \caption{A summary of previous work on the problem of the localized normal loading of an elastic half-space  that is coated by a thin layer (i.e.~$\ts/\tf\to\infty$).  The type of approach used in each reference is indicated by N (Numerics), E (Experiments) and/or A (Analysis).
  }
  \label{table:1}
\begin{tabular*}{\textwidth}{@{\extracolsep{\fill}}cccccc}
  \\
\textbf{Reference} &  \textbf{Modulus ratio} & \textbf{Load size} & \textbf{Load type} &  \textbf{Coating} & \textbf{Approach}\\
\hline\\
Yu \emph{et al.}~\cite{Yu1990} & $\Ef/\Es\leq 10$  &$\rind/\tf\geq 0.04$& Rigid indenter, various shapes & 3D solid & N \\
\\
Gao \emph{et al.}~\cite{Gao1992} &$1/2\leq\Ef/\Es\leq2$  & $1/7\leq \rind/\tf\leq 2$ & Rigid cylinder & 3D solid &N \& A\\
\\
Perriot \& Barthel \cite{Perriot2004} &$10^{-2}\leq\Ef/\Es\leq10^2$    & $10^{-3}\leq \rind/\tf\leq 10^2$ &Rigid indenter, various shapes& 3D solid & N\\
\\
Li \emph{et al.}~\cite{Li1997} &$10^{-1}\leq\Ef/\Es\leq10$    & $10^{-1}\leq \rind/\tf\leq 10$ & Parabolic pressure& 3D solid & N \\
\\
Liu \emph{et al.}~\cite{Liu2019} & $\Ef/\Es\gg1$    & $20\leq \rind/\tf\leq 10^3$ & Constant pressure & Beam & N \& E\\
\\
\hline\\
Current work & $\Ef/\Es\gg1$ &   $\tfrac{\rind}{\tf}\gg\left(\tfrac{\Es}{\Ef}\right)^{2/3}$ & Rigid cylinder & Beam & A, E \& N \\
\\
   \hline
  \end{tabular*} 
\end{table*}

\section{Scaling analysis\label{sec:scalings}}

We begin by noting that when subjected to a localized vertical displacement of size $\delta$, the coating would like the substrate to be deformed over a horizontal distance, $\lstar$, that is as large as possible, since this will minimize its curvature $\sim \delta/\lstar^2$ (and hence its bending energy $U_B\sim B\int (\delta/\lstar^2)^2~\upd A\sim B (\delta/\lstar)^2$). However, the elastic substrate opposes large $\lstar$: the typical strain $\varepsilon\sim \delta/\lstar$ is distributed over a volume $V\sim\lstar^3$ and so the substrate's elastic energy $\Ustrain\sim\Es\int\varepsilon^2~\upd V\sim \Es\delta^2\lstar$ increases with $\lstar$. Minimizing the total elastic energy $\Uelast=U_B+\Ustrain$ by varying $\lstar$, we find that the optimal horizontal length scale is $\lstar\sim (B/\Es)^{1/3}$. In the more detailed modelling that follows (see \S\ref{sec:MathModel}) it will be convenient to use the modified modulus $\Estar=\Es/(1-\nu_s^2)$ and to introduce an additional factor $2$ into our definition of $\lstar$; we therefore make the formal definition
\beq
\lstar=\left(\frac{2B}{\Estar}\right)^{1/3}.
\label{eqn:lstarDefn}
\eeq

Note that the plate model for the coating is only valid when $\lstar\gg \tf$, which we can see from \eqref{eqn:lstarDefn} requires
\beq
\Ef/\Es\gg1.
\eeq The analysis presented in this paper is therefore only valid for stiff coatings on soft substrates, as already anticipated in the introduction.

With the proviso that we are considering extremely large coating:substrate stiffness ratios, and assuming that the lateral deformation occurs over the energetically-optimal horizontal scale $\lstar$ given in \eqref{eqn:lstarDefn}, the total elastic energy of the system $\Uelast\sim \Es^{2/3}B^{1/3}\delta^2$. This energy must be provided to the system by the work of the indentation force $F$, which in scaling terms can be written $\Uindent\sim F\delta$. Hence, at a scaling level we expect that
\beq
F\sim \Es^{2/3}B^{1/3}\delta.
\label{eqn:Fscale}
\eeq Note that the bending of the coating leads to a constant indentation stiffness, $\kappa=F/\delta\sim \Es^{2/3}B^{1/3}$. The existence of a constant indentation stiffness is qualitatively similar to the Hertz contact result discussed in the introduction in which the coating alone is indented and, as such, would suggest a stiffness $\kappa\sim \Ef\rind$. Since we now have two estimates of the stiffness, the question then naturally arises of which of these best describes the stiffness that would be observed experimentally? The answer to this question depends on whether the coating deforms locally, as in Hertz contact, or rather bends, as in the argument that led to \eqref{eqn:Fscale} --- the softer of these two choices will be energetically favourable, and hence the expected mode of deformation. We therefore expect to observe the bending response \eqref{eqn:Fscale} when $\Es^{2/3}B^{1/3}\ll\Ef\rind$, or $\rind\gg \Es^{2/3}B^{1/3}/\Ef\sim\tf(\Es/\Ef)^{2/3}$: for sufficiently large indenters the coating will bend, deforming the substrate, rather than compress locally.

While the scaling law of \eqref{eqn:Fscale} is a useful first result, it relies on energy scalings that assumed a localized indenter. We shall see in \S\ref{sec:DetailedNums} that the assumption of a localized indenter is not necessarily at odds with the above calculation that bending deformation  occurs only for $\rind\gg \tf(\Es/\Ef)^{2/3}$ because $\Es/\Ef\ll1$. Nevertheless, the characteristic size of the indenter, $\rind$, does play a key role in determining the indentation stiffness, $\kappa=F/\delta$. To see why this should be the case, note that when $\rind\gg\lstar$ (a large indenter), the volume of the substrate that is strained by indentation is $\rind^3$ (rather than $\lstar^3$) and the elastic energy of indentation $\Ustrain\sim \Es\delta^2\rind$. This suggests that in this limit $F\sim \Es\rind\delta$ --- the constant Boussinesq indentation stiffness  for a cylindrical punch \cite{Sneddon1965} with the substrate (rather than coating) stiffness.  We therefore generalize \eqref{eqn:Fscale} to include a dependence on the dimensionless indenter size
\beq
\Rind=\frac{\rind}{\lstar}=\rind\left(\frac{\Estar}{2B}\right)^{1/3}
\label{eqn:EpsDefn}
\eeq 
by writing
\beq
\frac{F}{\delta}=\kappa ={\Estar}^{2/3}B^{1/3}\tkappa(\Rind).
\eeq

This paper is concerned primarily with the determination of the dimensionless stiffness $\tkappa(\Rind)$; we begin with an experimental determination of this function, presented in \S\ref{sec:ModelExpt}, before moving on to a theoretical calculation of $\tkappa(\Rind)$ in \S\ref{sec:MathModel} and comparing this to results obtained in related scenarios previously in \S\ref{sec:Previous}.

\section{Model experiments\label{sec:ModelExpt}}

Soft substrates with a thin, stiff coating were fabricated in the laboratory. The substrates were made from polyvinylsiloxane (PVS) elastomer (Elite Double 8, 22 and 32, Zhermack, Italy) by mixing a base polymer with a catalyst. The mixture was first degassed in a vacuum chamber and then cured within a cylindrical mould (with radius in the range $20\mathrm{~mm}\leq R_{\mathrm{subs}}\leq 55\mathrm{~mm}$ and substrate depth $15\mathrm{~mm}\leq \ts\leq 33\mathrm{~mm}$). The percentage of base polymer to catalyst in the elastomer mixture was varied to achieve  Young's moduli in the range $30\mathrm{~kPa} \leq \Es \leq 720\mathrm{~kPa}$ (see Appendix A); the  stiffness of uncoated substrates was measured by flat-punch indentation tests. The soft substrates were then coated with thin plastic films (RS Pro Shim Kit, RS Components Ltd., UK and Mylar, DuPont Teijin Films, US), of Young's Modulus $3.5\mathrm{~GPa} \leq \Ef \leq 5.7\mathrm{~GPa}$, and Poisson's ratio $\nu_f=0.4$. The thickness of the films $50\mathrm{~\mu m} \leq \tf \leq 128\mathrm{~\mu m}$, was measured optically using a microscope (Leica, DMIL, Leitz Wetzlar, Germany).  The films adhered to the soft substrate by contact alone; no additional adhesives were introduced into the system.  

\begin{figure}
 \centering
 \includegraphics[width=0.6\columnwidth]{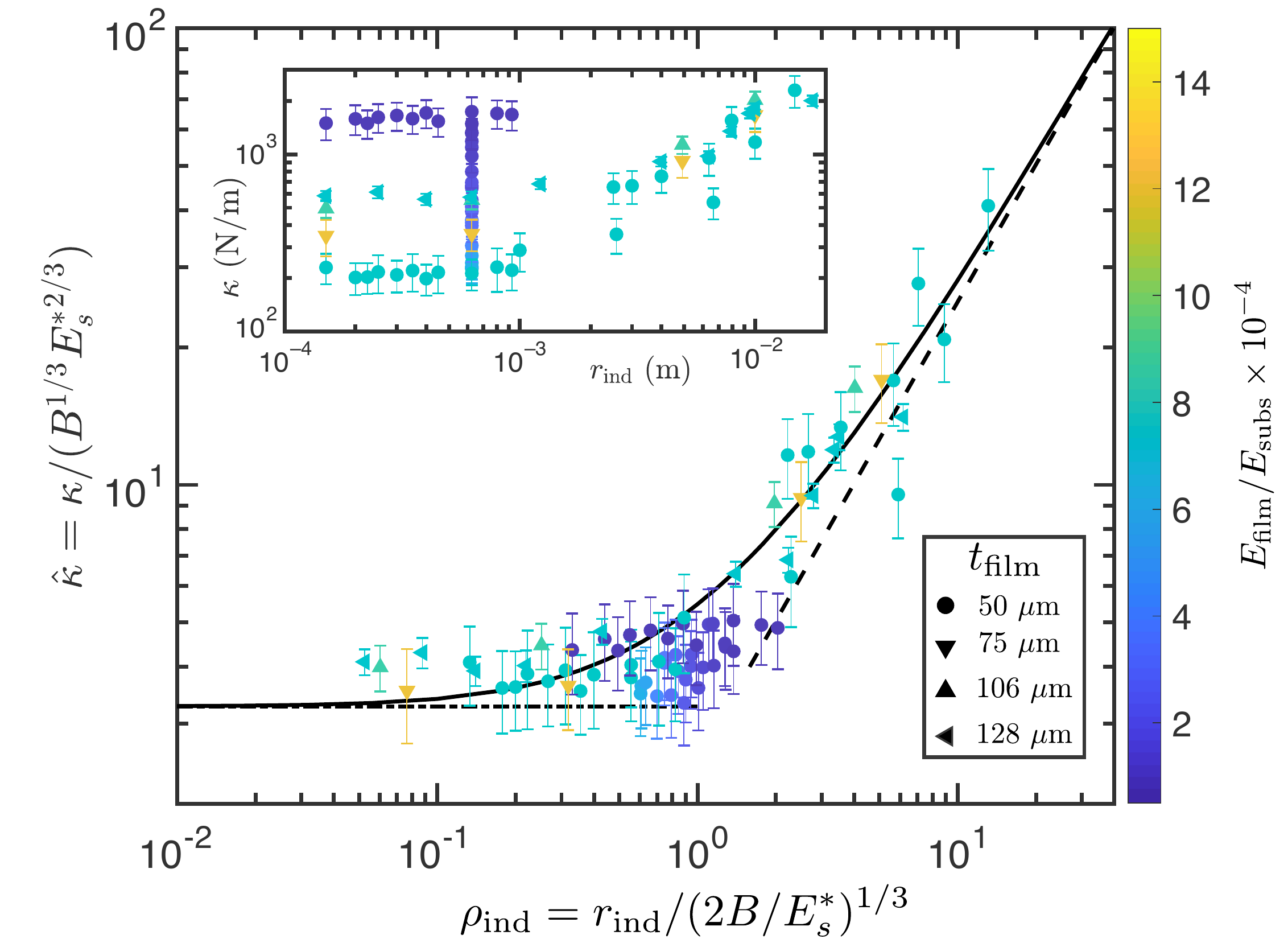}
 \caption{The results of model experiments. Inset: Raw measured values of $\kappa=F/\delta$ for different combinations of sheet and substrate as well as indenter radius, $\rind$. Main figure: Rescaled values of the experimentally measured stiffness $\tkappa$ (points) as a function of dimensionless indenter radius $\Rind=\rind/\lstar$. The theoretical prediction obtained from the numerical solution of the model developed here, described in \S\ref{sec:MathModel}, is shown by the solid curve, together with the asymptotic results for $\Rind\ll1$ (dash-dotted line) and $\Rind\gg1$ (dashed line), both reported in \eqref{eqn:NonDimK}. The elastic mismatch $\Ef/\Es$ is encoded by colour as indicated in the colour bar to the right, while the shape of the points shows the sheet thickness as shown in the legend.  }
 \label{fig:ModelExpts}
 \end{figure}

To measure the apparent stiffness of the resulting coated substrates, indentation tests were performed with flat-tipped, cylindrical indenters of different radius.  We therefore take the characteristic size of the indenters to be the cylinder radius, $\rind$, which was varied in the range $0.15\mathrm{~mm}\leq \rind \leq 17.6\mathrm{~mm}$. (These cylinders were stainless steel; for the narrowest cylinders, radii $\rind<1\mathrm{~mm}$,  syringe tips were used with the central hole filled with superglue to ensure contact throughout the tip region.) Typically, samples were positioned upon a microbalance (Pioneer, PA64C Analytic Balance, Ohaus, Switzerland), which measures forces accurate to within $0.1\mathrm{~mg}$, although larger samples were positioned upon a precision balance (PCB, 6000-0, Kern GmbH, Germany) with higher weighing capacity but lower precision (accurate to $1\mathrm{~g}$).  The centre of the sample was indented at $100\mathrm{~\mu m\,s^{-1}}$ using a linear actuator (M228, Physik Instrumente, Germany) controlled by a computer-controlled stepper motor (Mercury Step C663, Physik Instrumente) with typical unidirectional repeatability of $2\mathrm{~\mu m}$. 
The samples were subject to  indentation depths $\delta \leq 30\,\mu\,$m$ < \tf$, with the inequality $\delta\lesssim\tf$ ensuring that the effect of any stretching of the coating is smaller than that caused by bending of the coating \cite{Chopin2008}. The applied force, $F(\delta)$, was measured by recording the mass reported by the mass balance at $100\mathrm{~Hz}$. The reported indentation stiffness, $\kappa=F/\delta$, was acquired from the gradient of the measured linear response of force-displacement curves. A minimum of nine tests were performed on each sample; the reported stiffness is the mean value with  error bars representing the standard deviation of the measurements.  For these shallow indentations, the dimensions of the coated substrate were found to have no measurable influence on the measured stiffness, since $R_{\mathrm{subs}}$ and $\ts$ were both large in comparison to $\lstar$.

The results of our experiments are presented in fig.~\ref{fig:ModelExpts}. Raw measurements of the indentation stiffness as a function of the indenter radius are shown in the inset of fig.~\ref{fig:ModelExpts} and show that a variation of more than an order of magnitude in the measured stiffness may be obtained simply by varying the indenter radius or substrate stiffness. The main portion of fig.~\ref{fig:ModelExpts} shows that these raw data are well collapsed by plotting the dimensionless indentation stiffness, $\tkappa=\kappa/(B^{1/3}{\Estar}^{2/3})$, as a function of the dimensionless indenter radius $\Rind=\rind/\lstar$, defined in \eqref{eqn:EpsDefn}. To understand the behaviour of the dimensionless stiffness as a function of indenter size, i.e.~the curve $\tkappa(\Rind)$ shown in fig.~\ref{fig:ModelExpts}, we now present a mathematical model of indentation.

\section{Mathematical model\label{sec:MathModel}}

\subsection{Theoretical formulation}

We develop a mathematical model for the axisymmetric deflection, $\zeta(r)$, of the coating on top of the substrate in response to a cylindrical indenter of radius $\rind$ imposing a vertical displacement $\delta$. We shall model the coating as an elastic plate of bending stiffness $B=\Ef^\ast\tf^3/12$ that is subject to a vertical loading, $p(r;\delta)$ from the indenter as well as a deflection-induced response from the substrate, $Q(r;\delta)$. Using a flat cylindrical punch allows for the vertical loading from the indenter to always act only in $r<\rind$; the constant, known value of $\rind$ simplifies the problem somewhat compared to other shapes of indenter. Nevertheless, both $p(r;\delta)$ and $Q(r;\delta)$ are \emph{a priori} unknown, so that they must be determined as part of the solution.

Neglecting any tension within the coating, the plate equation\cite{Ventsel2001} for the vertical deflection of the coating with a specified indentation depth $\delta$  reads
\beq
B\nabla^4\zeta=p(r;\delta)+Q(r;\delta).
\label{eqn:plateeqn}
\eeq This is to be solved with the conditions
\beq
\zeta(r)=-\delta,\quad r<\rind
\eeq as well as far-field conditions $\zeta,\upd\zeta/\upd r\to0$ as $r\to\infty$.

To make analytical progress in determining the response of the elastic substrate, $Q(r;\delta)$, to a vertical deflection $\zeta(r;\delta)$, we make use of Hankel transforms --- the Hankel transform of a function $f(r)$ is $\hankel{f}(k)=\int_0^\infty rf(r)J_0(kr)~\upd r$ where $J_0(x)$ is the zeroth-order Bessel function \cite{Abramowitz1964} and $k$ is the scaling factor, analogous to wave number or frequency in a Fourier transform. A classic result of Sneddon \cite{Sneddon1995} is that, for a substrate of infinite depth, and neglecting any shear stress on the top surface of the substrate (i.e.~neglecting the effect of a tension within the elastic sheet on the substrate) the Hankel transform of the normal load on the sheet from the substrate, $\hankel{Q}(k)$, is proportional to the Hankel transform of the interfacial deflection, $\hankel{\zeta}(k)$. In particular, Sneddon~\cite{Sneddon1995} showed that
\beq
\hankel{Q}(k)=-\tfrac{1}{2}\Estar k\hankel{\zeta}(k).
\label{eqn:SneddonQ}
\eeq  (Note that for an incompressible substrate, $\nu_s=1/2$, the solution leading to \eqref{eqn:SneddonQ} has both zero shear stress at the surface of the soft substrate and zero horizontal displacement \cite{Sneddon1995}.)

\subsection{Solution of the problem}

Substituting the expression from \eqref{eqn:SneddonQ} into the Hankel transform of \eqref{eqn:plateeqn} we find that
\beq
k\left[Bk^3+\tfrac{1}{2}\Estar\right]\hankel{\zeta}(k)=\hankel{p}(k).
\label{eqn:zetatilde}
\eeq
We therefore have an explicit expression for the Hankel transform of the pressure applied by the indenter in terms of the Hankel transform of the vertical displacement of the coating everywhere. To proceed further we  make use of the facts that: (i) $p(r)$ vanishes for $r>\rind$ (since this is beyond the indenter) and (ii) $\zeta(r)=-\delta$ for $r<\rind$ (since this is within the region displaced by the cylindrical indenter). These conditions may be written in terms of the inverse Hankel transforms of $\hankel{p}(k)$ and $\hankel{\zeta}(k)$ as:
\beq
p(r)=\int_0^\infty k\hankel{p}(k)J_0(kr)~\upd k=0,\quad r>\rind
\label{eqn:ptilde1}
\eeq and
\beq
\zeta(r)=\int_0^\infty k\hankel{\zeta}(k)J_0(kr)~\upd k=-\delta,\quad r<\rind,
\label{eqn:ztilde1}
\eeq respectively.

Using \eqref{eqn:zetatilde}, $\hankel{\zeta}(k)$ may be eliminated from \eqref{eqn:ztilde1} in favour of $\hankel{p}(k)$, which leads to
\beq
\int_0^\infty \frac{\hankel{p}(k)}{Bk^3+\Estar/2}J_0(kr)~\upd k=-\delta,\quad r<\rind.
\label{eqn:ptilde2}
\eeq

Equations \eqref{eqn:ptilde1} and \eqref{eqn:ptilde2} are a pair of integral equations, which can, in principle, be solved to determine the Hankel transform of the indenter pressure, $\hankel{p}(k)$. We shall, in general, have to perform this inversion numerically. However, before analysing these equations further, we first  consider the behaviour in two asymptotic limits that can be solved analytically.

\subsubsection{A point indenter: $\rind\to0$}

For a point indenter,  the indentation pressure
\beq
p(r)=\frac{F}{2\pi}\frac{\delta(r)}{r},
\eeq which is obtained as the limit of an indentation force $F$ uniformly distributed over a small circle in the limit of vanishing circle radius  \cite{Ventsel2001,Taffetani2017}. We therefore have  $\hankel{p}(k)=F/(2\pi)$ so that, using \eqref{eqn:ptilde2}, we find
\beq
-\delta=\zeta(0)=\frac{F}{2\pi}\int_0^\infty\left[Bk^3+\tfrac{1}{2}\Estar\right]^{-1}\mathrm{~\upd}k.
\eeq After computation of the integral $\int_0^\infty(X^3+1)^{-1}~\upd X=2\pi/3^{3/2}$, we then have that $F=-\kappa_0\delta$ where the indentation stiffness of a point-like indenter is
\beq
\kappa_0=\frac{3^{3/2}}{2^{2/3}}B^{1/3}{\Estar}^{2/3}.
\label{eqn:kPoint}
\eeq Note that in the limit of a point indenter, therefore, the apparent stiffness of the combined material mixes the substrate stiffness $\Estar$ with the bending stiffness of the coating $B$ in the manner expected from the scaling analysis of \S\ref{sec:scalings}. However, we have now also been able to determine the appropriate pre-factor.

Knowledge of the Hankel transform of the pressure in this limit allows us to use \eqref{eqn:ztilde1} to show that
\beq
\zeta(r)=\frac{F}{2\pi}\int_0^\infty\frac{J_0(kr)}{Bk^3+\tfrac{1}{2}\Estar}\mathrm{~\upd}k.
\eeq Changing variable to $K=kr$, we find that for $r\gg\lstar$
\beq
\zeta(r)\approx \frac{F}{\pi\Estar r}.
\label{eqn:zetaFar}
\eeq  This  far-field behaviour exhibits algebraic decay, $\zeta(r)\sim r^{-1}$ as $r\to\infty$, explaining the need for relatively large substrates in experiments. Moreover, we expect similar results to hold far from other sized indenters since the substrate feels only the total applied force, $F$, far from the indenter. To understand the coating deflection in the vicinity of the indenter's edge, it may be possible to follow a boundary layer analysis of the type presented for a similar problem with surface tension \cite{Karpitschka2016}; we do not investigate this possibility here since our focus lies in the force--displacement relationship.

\subsubsection{No coating: $B\to0$}

In the limit $B\to0$, we expect to recover the classic result for an uncoated substrate due to Sneddon \cite{Sneddon1965} amongst others. In particular, letting $B=0$ in the integral equations \eqref{eqn:ptilde1} and \eqref{eqn:ptilde2}, we find a system that is solved precisely by Sneddon's solution\cite{Sneddon1965}, i.e.  
\beq
\hankel{p}(k)=-\delta\frac{\Estar}{\pi}\frac{\sin^{-1}(k\rind )}{k},
\eeq which corresponds to
\beq
p(r;\delta)=-\delta\frac{\Estar}{\pi}(\rind^2-r^2)^{-1/2},\quad r<\rind.
\label{eqn:SneddonPressure}
\eeq The indentation force can then be calculated as
\beq
F=2\pi\int_0^{\rind}rp(r;\delta)~\upd r=-2\Estar\rind\delta
\label{eqn:IndForce}
\eeq so that the quantity of most interest to us here, the indentation stiffness in the uncoated limit,  is simply
\beq
\kappa_\infty=\left.\frac{-F}{\delta}\right|_{B=0}=2\Estar\rind.
\label{eqn:SneddonStiffness}
\eeq (This is precisely the solution of the Boussinesq problem discussed in the Introduction.)

We shall see that the limit $B=0$ is equivalent to that of a sufficiently large indenter, $\rind/\lstar\gg1$ and, further, that this limit is well-defined (and non-singular). To see this, we now discuss the non-dimensionalization of the problem.

\subsubsection{Non-dimensionalization}

There are two natural length scales in the problem: the indenter size $\rind$ and the coating--substrate length scale $\lstar=(2B/\Estar)^{1/3}$. We shall use the indenter radius $\rind$ as the natural length scale, introducing the dimensionless parameter $\Rind$ given in \eqref{eqn:EpsDefn}. It is also clear from the linearity of \eqref{eqn:ptilde1} and \eqref{eqn:ptilde2} that the applied pressure is linear in the indentation depth $\delta$. We therefore non-dimensionalize the problem by letting
\beq
R=r/\rind,\quad K=k\rind, \quad P=p/(B\delta/\rind^4).
\eeq The integral equations \eqref{eqn:ptilde1} and \eqref{eqn:ptilde2} then become
\beq
\int_0^\infty K\hankel{P}(K)J_0(KR)~\upd K=0,\quad R>1
\label{eqn:ptildeND1}
\eeq and
\beq
\int_0^\infty \frac{\hankel{P}(K)}{K^3+\Rind^3}J_0(KR)~\upd K=-1,\quad R<1
\label{eqn:ptildeND2}
\eeq respectively. (Note that the factor of $2^{1/3}$ in the earlier choice of $\lstar$ was included to simplify the denominator in the integrand of \eqref{eqn:ptildeND2}.)

The dimensionless indentation stiffness can then be determined from the force condition \eqref{eqn:IndForce} to be
\beq
\tkappa=\frac{\kappa}{B^{1/3}{\Estar}^{2/3}}=2\pi\Rind^{-2}\int_0^1RP(R)~\upd R.
\label{eqn:KFromNums}
\eeq

Having non-dimensionalized the problem, we now see that the earlier analytical results may be written in dimensionless form as
\beq
\tkappa=\frac{\kappa}{B^{1/3}{\Estar}^{2/3}}\sim\begin{cases}
\frac{3^{3/2}}{2^{2/3}},\quad \Rind\ll1\\
2^{4/3}\Rind,\quad \Rind\gg1. 
\end{cases}
\label{eqn:NonDimK}
\eeq However, we would like to have results for a wider range of values of $\Rind$. This requires a numerical solution of the integral equations \eqref{eqn:ptildeND1}--\eqref{eqn:ptildeND2}, and so we  turn to discuss this problem next.

\subsection{Numerical results}

\begin{figure}
 \centering
 \includegraphics[width=0.78\columnwidth]{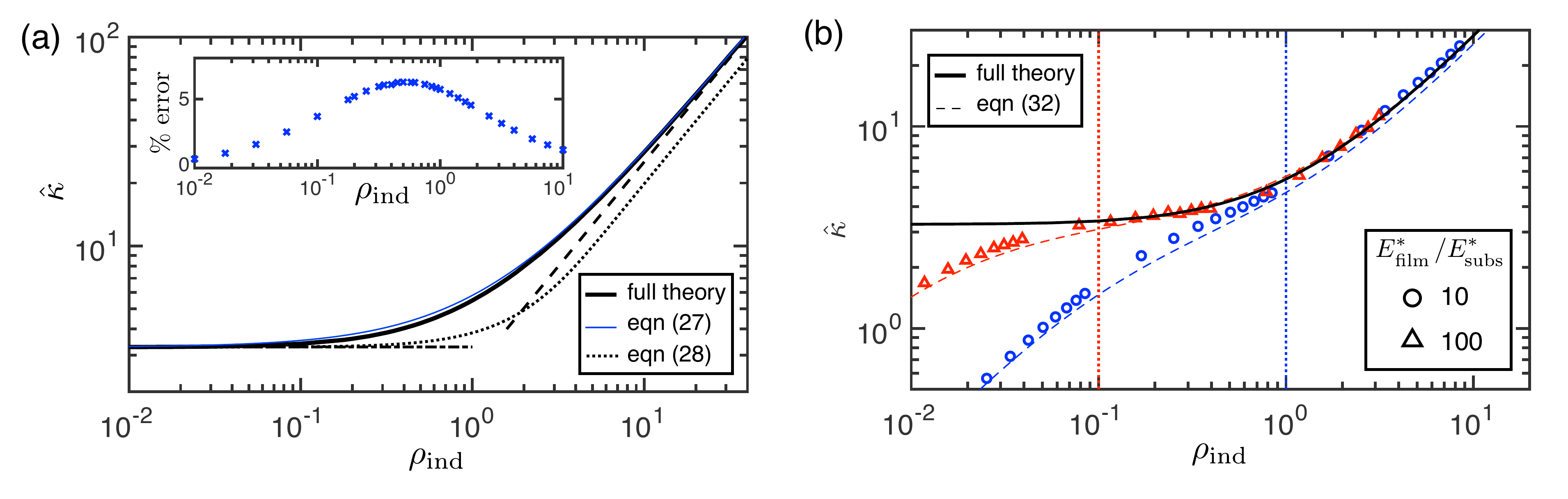}
 \caption{The  dimensionless indentation stiffness $\tkappa=\kappa/(B^{1/3}{\Estar}^{2/3})$ as a function of the dimensionless indenter radius $\Rind=\rind/\lstar$, determined from the numerical solution of our model. (a) Main figure: Numerical results for $\tkappa(\Rind)$ (thick solid curve) compared with a variety of approximate expressions: the approximation \eqref{eqn:Kapprox} is shown by the thin solid blue curve while the approximate result of ref.~\cite{Liu2019}, \eqref{eqn:LiuK}, is shown as the dotted curve. Asymptotic results from \eqref{eqn:NonDimK} are shown for $\Rind\ll1$ (dash-dotted line) and $\Rind\gg1$ (dashed line). Inset: The error introduced by using the approximate formula \eqref{eqn:Kapprox} rather than the numerical solution of the system of equations \eqref{eqn:ptildeND1}--\eqref{eqn:ptildeND2}. (b)  Comparison between previous numerical results of the indentation of a coated substrate by a cylindrical punch (points) and the numerical solution of our reduced model (solid curve). The numerical results for the coated problem are reproduced from a digitization of figure 2 of ref.~\cite{Perriot2004} with stiffness ratios $\Ef^\ast/\Estar=10$ (circles) and $\Ef^\ast/\Estar=100$ (triangles). Dotted vertical lines show where $\Rind=10\Estar/\Ef^\ast$ for each value of $\Ef^\ast/\Estar$; we expect our model of the coating as an elastic plate to be valid only for $\Rind\gg\Estar/\Ef^\ast$, as in \eqref{eqn:HertzLimit}. Dashed curves show the approximate relationship \eqref{eqn:HertzApprox} with the appropriate value of $\Estar/\Ef^\ast$; this combines the Hertzian behaviour of the coating with the plate bending response of the coated substrate.}
 \label{fig:MokhtarNums}
 \end{figure}

We solve the  integral equations \eqref{eqn:ptildeND1}--\eqref{eqn:ptildeND2} numerically, as detailed in Appendix B. Our results allow the  stiffness $\kappa$ to be determined for various values of the dimensionless indenter size $\Rind$. The behaviour of the dimensionless stiffness $\tkappa=\kappa/(B^{1/3}{\Estar}^{2/3})$ as the dimensionless indenter size $\Rind=\rind/\lstar$ varies is plotted with the experimental data in fig.~\ref{fig:ModelExpts} and shows good agreement between the two. These numerical results are also plotted in fig.~\ref{fig:MokhtarNums}, and illustrate that the asymptotic results of \eqref{eqn:NonDimK} are well reproduced by the  numerical solution of our model in the appropriate limits. However, we also note that the approximate expression
\beq
\tkappa(\Rind)\approx \frac{3^{3/2}}{2^{2/3}}+2^{4/3}\Rind, 
\label{eqn:Kapprox}
\eeq determined by adding the two asymptotic expressions, agrees with  the numerical solution of our model  to within $6.3\%$ across all values of $\Rind$ (see inset of fig.~\ref{fig:MokhtarNums}a).

While the agreement between the numerical solution of the model presented here and the various asymptotic results presented in fig.~\ref{fig:MokhtarNums} is very good, these results depend on a series of modelling assumptions, most notably that the thin coating may be modelled as an elastic plate. The good agreement with the experiments presented in fig.~\ref{fig:ModelExpts} suggests that this approximation is appropriate in this case. However, to offer a more stringent test of this modelling assumption, we also consider how the results we have presented here compare with previous  results on related problems.

\section{Comparison with previous results\label{sec:Previous}}

\subsection{Theoretical results}

Liu \emph{et al.}~\cite{Liu2019} followed  a theoretical approach very similar to that adopted here. However, rather than solving the pair of integral equations \eqref{eqn:ptildeND1}--\eqref{eqn:ptildeND2}, they assume that the pressure distribution is uniform in the contact region (vanishing beyond this region), i.e.~$p(r)=F/(\pi\rind^2)$ in $r<\rind$. This uniform pressure distribution has Hankel transform $\tilde{p}(k)=F J_1(k\rind)/(\pi k\rind)$; this does identically satisfy the first equation \eqref{eqn:ptildeND1} (by assumption) but is inconsistent with a constant vertical displacement within the contact region, $r<\rind$. Nevertheless, an estimate of the indentation stiffness may be determined by evaluating \eqref{eqn:ptilde2} at $r=0$. In our notation, this approximation reads:
\beq
\tkappa(\Rind)\approx\frac{\pi}{2^{2/3}}\Rind\left[\int_0^\infty\frac{J_1(K\Rind)}{K(K^3+1)}~\upd K\right]^{-1},
\label{eqn:LiuK}
\eeq  which is easily computed numerically (or indeed analytically, albeit in terms of the Meijer G-function\cite{Olver2010}). Moreover, for $\Rind\ll1$ the expression \eqref{eqn:LiuK} reproduces the appropriate asymptotic result from \eqref{eqn:NonDimK} --- for small indenters, the deviation from uniform vertical displacement beneath the indenter is only small. However, for $\Rind\gg1$ we find that $\tkappa\sim \pi\Rind/2^{2/3}$, which yields a systematic, and constant, error of more than $20\%$ compared to the large indenter limit of \eqref{eqn:NonDimK}, $\tkappa\sim2^{4/3}\Rind$. The comparison of this result with the numerical solution of our model shown in fig.~\ref{fig:MokhtarNums}a demonstrates that the values at intermediate $\Rind$ are also very different: \eqref{eqn:LiuK} consistently underestimates the  stiffness determined from our model calculation. This discrepancy reflects the fact that the pressure is far from uniform for non-small indenters (for example, Sneddon \cite{Sneddon1965} showed that the pressure actually diverges at the edge of a cylindrical indenter, as shown in \eqref{eqn:SneddonPressure}).

\subsection{Detailed numerical results\label{sec:DetailedNums}}

The results we have presented rely on modelling the deformation of the thin coating as an elastic plate. To test the validity of this simplifying approximation,  we compare our results with numerical results reported by Perriot \& Barthel \cite{Perriot2004} that lift this restriction.  In particular, numerical results from fig.~2 of ref.~\cite{Perriot2004} were captured digitally and are plotted in  figure \ref{fig:MokhtarNums}b (after translating to the non-dimensionalization of  the present paper). When plotted in the way suggested by our theory, these results collapse with the collapse being particularly good at large values of the indenter radius, $\Rind\gtrsim1$. However, at smaller indenter radius, both the collapse and the agreement with  the results of our model, specifically the point indenter limit \eqref{eqn:kPoint}, break down.

To understand the discrepancy between the numerical results presented by Perriot \& Barthel \cite{Perriot2004} and the results of the model presented here, we revisit our assumption that when the indenter contacts the substrate, the whole coating bends beneath it. An alternative mode of deformation is that the coating itself compresses, which is frequently referred to as Hertz contact\cite{Johnson1987}. This mode of deformation has typical indentation stiffness $k_{\mathrm{Hertz}}^{\mathrm{film}}\sim \Ef^\ast\rind$. For sufficiently small indenters, this deformation mode may be `softer' than the  bending deformation we have considered, which had stiffness $k_{\mathrm{bend}}^{\mathrm{film}}\sim B_f^{1/3}{\Estar}^{2/3}\sim \tf{\Ef^\ast}^{1/3}{\Estar}^{2/3}$. We may consider these different modes of deformation to be linear springs acting in series, and so expect the measured stiffness to be dominated by whichever is the softer; in particular, we expect to observe the bending response studied in this paper provided it is `softer' than the Hertz-like response of the coating, i.e.
\beq
1\gg\frac{k_{\mathrm{bend}}^{\mathrm{film}}}{k_{\mathrm{Hertz}}^{\mathrm{film}}}\sim \frac{ \tf{\Ef^\ast}^{1/3}{\Estar}^{2/3}}{\Ef^\ast\rind}\sim\frac{\tf}{\rind}\left(\frac{\Estar}{\Ef^\ast}\right)^{2/3}
\eeq which in turn requires that
\beq
\frac{\rind}{\tf}\gg\left(\frac{\Estar}{\Ef^\ast}\right)^{2/3},
\label{eqn:plateDefn}
\eeq or
\beq
\Rind=\frac{\rind}{\lstar}\gg \frac{\Estar}{\Ef^\ast}.
\label{eqn:HertzLimit}
\eeq Vertical lines corresponding to $\Rind=10\Estar/\Ef^\ast$ are shown in fig.~\ref{fig:MokhtarNums}b, and approximately coincide with the values of $\Rind$ at which the disagreement between the numerical solutions of ref.~\cite{Perriot2004} and the numerical solution of our own theoretical model is noticeable.

To be more quantitative, we take the analogy of springs in series further: a given imposed force $F$ will induce a displacement caused by the bending of the coating \emph{and} a displacement caused by the localized (Hertzian) compression of the coating itself. Adding these two displacements gives the total displacement caused by the force $F$, $\delta_{\mathrm{total}}\approx F\bigl\{(2\Ef^{\ast}\rind)^{-1}+\bigl[B^{1/3}{\Estar}^{2/3}\tkappa(\Rind)\bigr]^{-1}\bigr\}$. If we approximate $\tkappa(\Rind)$ using \eqref{eqn:Kapprox}, we readily find a combined stiffness $\kappa_{\mathrm{comb}}=B^{1/3}{\Estar}^{2/3}\tkappa_{\mathrm{comb}}$ with
\beq
\tkappa_{\mathrm{comb}}\approx\left(\frac{3^{2/3}}{2^{2/3}}+2^{4/3}\Rind\right)\left[1+\frac{\Es^\ast}{\Ef^\ast}\left(1+\frac{3^{3/2}}{4\Rind}\right)\right]^{-1}.
\label{eqn:HertzApprox}
\eeq The comparison between the numerical results of ref.~\cite{Perriot2004} and \eqref{eqn:HertzApprox} is shown in fig.~\ref{fig:MokhtarNums}b; for $\Ef^\ast/\Es^\ast=10$ and  $\Ef^\ast/\Es^\ast=100$, the maximum relative error of \eqref{eqn:HertzApprox} across all indenter sizes is $13.9\%$ and $9.6\%$, respectively.

We note that the condition of \eqref{eqn:HertzLimit} holds for the experiments we presented in \S\ref{sec:ModelExpt}, since for our experiments $\Estar/\Ef^\ast\lesssim10^{-3}$ while the dimensionless indenter radius $\Rind\gtrsim10^{-1}$. Finally, we note that the requirement of \eqref{eqn:HertzLimit} may be compatible with the point indenter limit, $\Rind=\rind/\lstar\ll1$, provided that $\Estar/\Ef^\ast\lll1$. In particular, for the point indenter limit of our model to be valid, we require $\Rind \ll1$ whilst simultaneously satisfying \eqref{eqn:HertzLimit}, i.e.~
\beq
\Estar/\Ef^\ast\ll\Rind \ll1.
\label{eqn:PtIndEps}
\eeq Alternatively, one may write the condition for a point indentation of a bending plate, \eqref{eqn:PtIndEps}, in terms of the ratio of the radius to the film thickness, which reads
\beq
\left(\frac{\Estar}{\Ef^\ast}\right)^{2/3}\ll \frac{\rind}{\tf}\ll\left(\frac{\Ef^\ast}{\Estar}\right)^{1/3}.
\eeq

\section{Discussion\label{sec:Discuss}}

\subsection{Summary of results\label{sec:summary}}

We have considered in detail the problem of small indentations of a soft substrate that is coated by a thin stiff layer.  We developed a model that combined  plate theory (to describe the deflection of the coating) with classic results for the deformation of a substrate due to an applied pressure distribution.  
By comparison with previous numerical results, we  showed that the plate model of the coating is valid provided that the substrate stiffness is significantly lower than that of the film; in particular from \eqref{eqn:plateDefn} we require indenter to thickness ratios $\rind/\tf\gg(\Estar/\Ef^\ast)^{2/3}$. Under this condition, and provided that the indentation depth remains small enough to neglect stretching within the coating ($\delta\lesssim\tf$), we find that the indentation `stiffness' depends on the indenter size, $\rind$.  In particular, for sufficiently small indenters, the indentation stiffness mixes the bending stiffness of the coating with the stiffness of the underlying substrate, while for sufficiently large indenters it is the substrate stiffness alone that determines the indentation stiffness.

Detailed asymptotic results are summarized in dimensionless terms in \eqref{eqn:NonDimK}, but may be rewritten in dimensional terms as:
\beq
\kappa=\frac{F}{\delta}\sim\begin{cases}
\frac{3^{3/2}}{2^{2/3}}B^{1/3}{\Estar}^{2/3},\quad \rind\ll\lstar\\
2\Estar\rind,\quad \rind\gg\lstar.
\end{cases}
\label{eqn:KDim}
\eeq  Note, in particular, that for small indenters the indentation stiffness measured relative to that of the uncoated substrate $\kappa/(2\Estar \rind)\propto \lstar/\rind\gg1$: for small indenters, the coating greatly stiffens the substrate, effectively cloaking its true modulus (see fig.~\ref{fig:Krelative}).

We have used the numerical solution of our model equations to determine the stiffness $\kappa$ for intermediate indenter radii, $\rind=O(\lstar)$. These results showed that a simple approximation valid throughout the range of indenter sizes may be obtained by adding the asymptotic limits --- the dimensionless expression in \eqref{eqn:Kapprox} is always within $6.3\%$ of the value determined from the numerical solution of our model. As a result, we suggest the dimensional version of \eqref{eqn:Kapprox}, namely
\beq
\kappa\approx\tfrac{3^{3/2}}{2^{2/3}}B^{1/3}{\Estar}^{2/3}+2\Estar\rind,
\label{eqn:KapproxDim}
\eeq may be used to predict the stiffness that would be measured with a particular indenter radius and known material properties. 

Typically, however, one uses indentation to determine the  unknown properties of a system, with little or no knowledge of the underlying material properties. Our results suggest that to determine both the bending stiffness of the coating and the Young's modulus of the substrate requires a suite of experiments with different indenter radii and repeated measurements of $\kappa$. By comparing a linear fit of $\kappa(\rind)$ with \eqref{eqn:KapproxDim} we see that the substrate Young's modulus should be half the linear slope while the bending stiffness of the coating can be inferred from the intercept as $\rind\to0$. However, once such a fit has been performed, one must also check that the indenters used were sufficiently large that our use of plate theory is satisfied, i.e.~that \eqref{eqn:plateDefn} is satisfied with the obtained parameter values (which requires, in addition, knowledge of the coating thickness).

\subsection{Relevance to previous experiments on fruit \label{sec:apples}}

We motivated our study of the indentation of soft substrates coated by a thin, stiff layer with the question of how one determines whether a piece of fruit is ripe (or not) without damaging it. We now turn again to this question to consider the insights that the analysis presented in the main body of this paper, and the results discussed in \S\ref{sec:summary}, in particular, might bring.

The first question is whether the various assumptions made in our analysis hold? In particular, is our use of plate theory to model the coating appropriate in this scenario? Apples seem to be the fruit with the most comprehensive set of published experimental data for comparison. Previous work by Grotte \emph{et al.}~\cite{Grotte2001} gives a typical modulus for the flesh of $\Es\approx500\mathrm{~kPa}$  while Wang \emph{et al.}~\cite{Wang2017} reported the skin to have typical modulus $\Ef\approx 20\mathrm{~MPa}$  and thickness $\tf\approx215\mathrm{~\mu m}$. These values give an estimate of $\lstar\approx400\mathrm{~\mu m}$. As a result, we expect that our plate model of the skin should be valid provided that
\beq
\rind\gg t\left(\frac{\Estar}{\Ef^\ast}\right)^{2/3}\approx 20\mathrm{~\mu m}.
\label{eqn:critEpsApple}
\eeq 

\begin{figure}
 \centering
 \includegraphics[width=0.6\columnwidth]{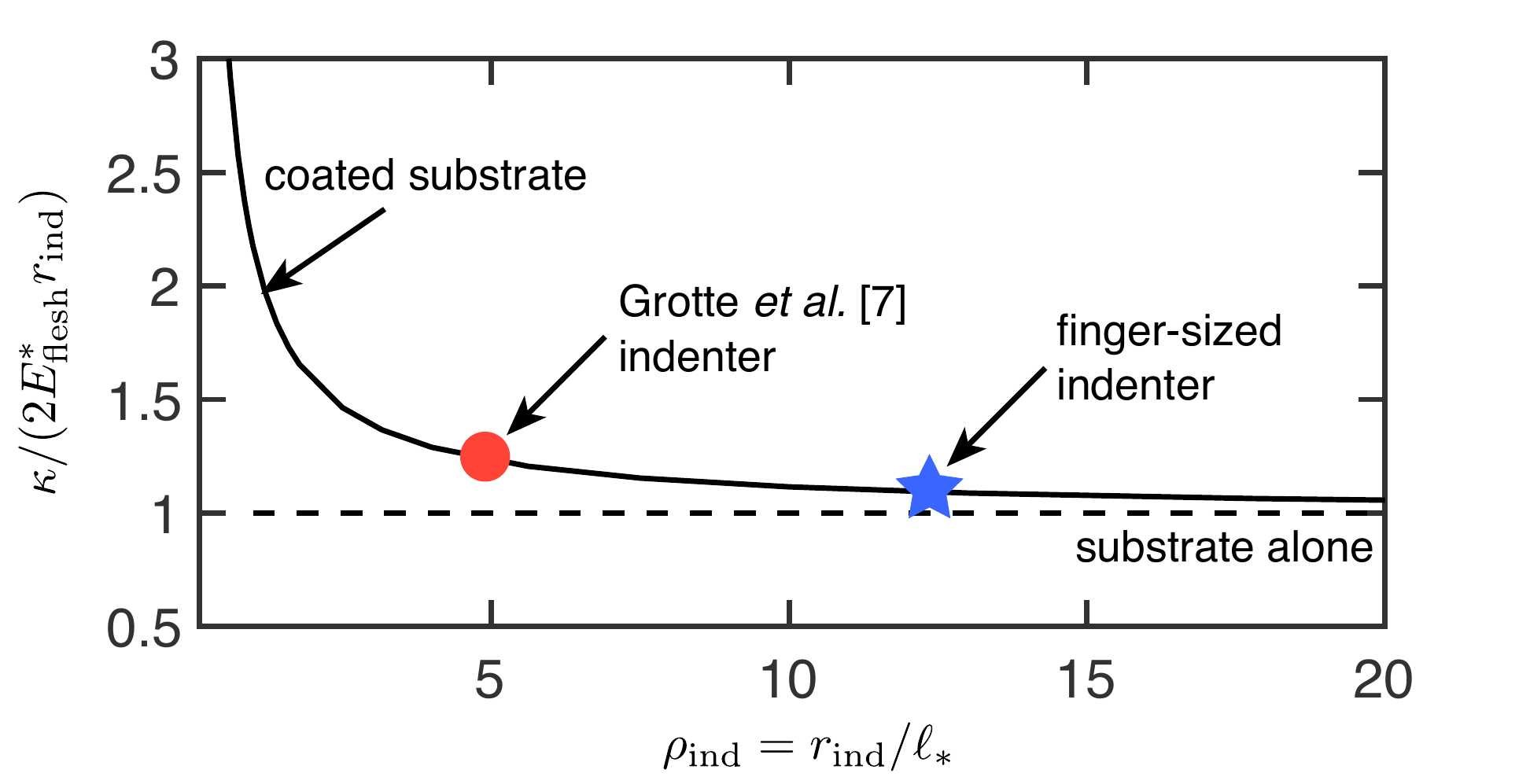}
 \caption{The effectiveness of `cloaking by coating': the indentation stiffness of a coated substrate measured  relative to the indentation stiffness of the uncoated substrate. Here, the solid curve shows the prediction based on the numerical solution of our model, while the dashed line shows the pure uncoated stiffness. The extent of cloaking by coating expected for an apple are shown by points: for the indenters typically used in industrial measures \cite{Grotte2001} of apple ripeness, $\rind=2\mathrm{~mm}$ (indicated by the circular point),   the presence of a stiffer skin means that the apparent stiffness is around $25\%$ larger than that of the underlying flesh; larger indenters, such as a finger (indicated by the star), give an indentation stiffness  less than $10\%$ above that of the underlying flesh. }
 \label{fig:Krelative}
 \end{figure}

Figure 2 of ref.~\cite{Grotte2001} presents indentation tests of an apple with and without the skin using a cylindrical indenter with diameter $2\rind=4\mathrm{~mm}$; such an indenter easily satisfies the condition \eqref{eqn:critEpsApple} under which we expect the plate theory approximation used here to be valid.  The experimental  results presented by Grotte \emph{et al.}~\cite{Grotte2001} show that with the skin intact, the measured `firmness' (our indentation stiffness) is increased by a factor of around 3 compared to situations in which the skin is first removed. This is significantly larger than the size of effect expected based on the theory presented here, which would predict that the skin should lead to an increase of around $25\%$ (see the circular point in fig.~\ref{fig:Krelative}). We discuss possible reasons for this discrepancy in the conclusion, but note also that a larger indenter (such as a finger) would yield a firmness within $10\%$ of that of the substrate itself (see star in fig.~\ref{fig:Krelative}).

\section{Conclusion}

We have presented a theoretical model for the increase in firmness that is provided by a stiff, thin coating of a soft substrate. This model, and its numerical solution, demonstrated the critical role of the indenter size in determining whether the coating significantly stiffens the substrate or not: loosely speaking, small indenters `feel' the effect of the coating, while large indenters feel the underlying substrate. 

The predictions of our theoretical model are in good agreement with model experiments on soft substrates coated by significantly stiffer thin films, and previously published detailed numerical simulations. However, our predictions seem to significantly under-estimate the effect of the skin-induced stiffening of fruit. We believe that this is likely due to the effect of a pre-existing tension within the skin, which resists indentation more effectively than the bending stiffness accounted for here. (The likely presence of such a pre-tension could be shown by introducing an incision in the skin and observing that the relaxation of the pre-tension leads to the spontaneous opening of the incision.) Another effect that might also be included in the modelling of this indentation process is the natural curvature of most fruit (though we do not expect this to be a significant effect for the apples presented in \S\ref{sec:apples} since the radius of an apple is significantly larger than the typical length scale $\lstar\approx400\mathrm{~\mu m}$).

\section*{Acknowledgements}

The research leading to these results has received funding from the European Research Council under the European Union's Horizon 2020 Programme / ERC Grant Agreement no.~637334 (DV) and the Leverhulme Trust through a Philip Leverhulme Prize (DV). We are grateful to Tom Chandler for discussions during the course of this work.

\section*{Appendix A: Obtaining different substrate Young's moduli}

\begin{figure}
 \centering
 \includegraphics[width=0.6\columnwidth]{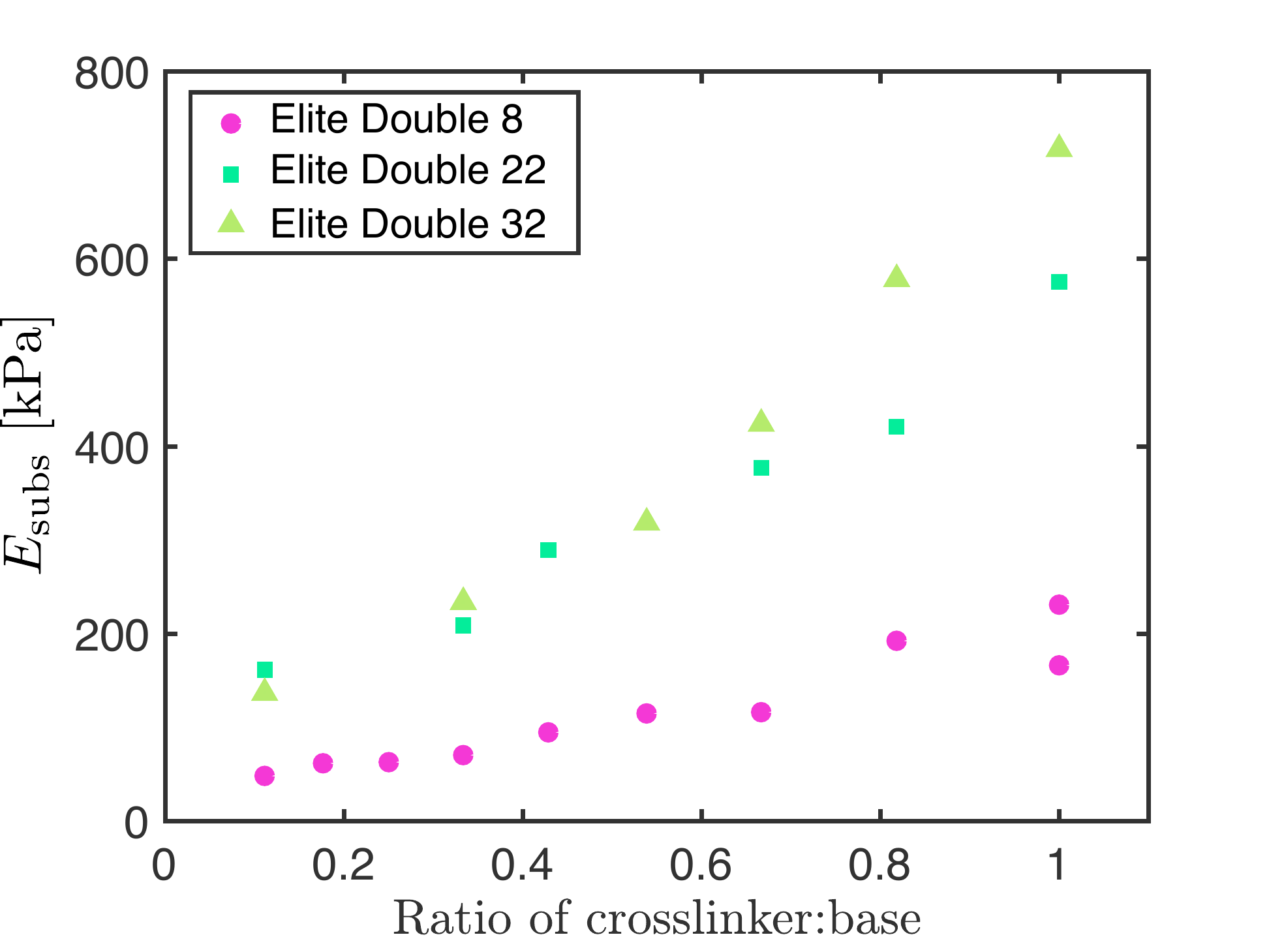}
 \caption{The measured value of the Young's modulus of the substrate, $\Es$, as a function of the ratio of crosslinker to base polymer. Results are shown for three different  grades of Polyvinylsiloxane (as indicated in the legend). The Young's modulus was measured using an indentation test with an uncoated, deep substrate for indentation depths $\delta<100\mathrm{~\mu m}$ and the measured indentation stiffness converted to a Young's modulus via Sneddon's result, \eqref{eqn:SneddonStiffness}, with $\nu_s=0.5$. The variance between repeated measurements of the same sample is less than $10\%$.}
 \label{fig:SubStiff}
 \end{figure}

Polyvinylsiloxane (PVS) is an elastomer that is fabricated by mixing a base polymer with a curing agent  (i.e.~a crosslinker). Ordinarily, the two parts are mixed in equal measures and the mixture allowed to set. However, it is well known that the mechanical properties of other elastomers, including Polydimethylsiloxane (PDMS), can be tuned by varying the degree of crosslinking in the polymer network \cite{Wang2014}. In the experiments presented here,  the stiffness of the PVS substrates  was varied by using mixtures with different amounts of crosslinker to each part of the  polymer base (reported as a ratio < 1 in fig.~\ref{fig:SubStiff} since all mixtures were at least 50\% base, with the softest corresponding to 90\% base). These different mixtures were fabricated for each of  three different grades of PVS (Elite Double 8, 22 and 32), supplied by Zhermack (Italy). The mixtures were thoroughly mixed, degassed in a vacuum chamber and left to cure in a cylindrical mould for one hour before  the mechanical properties were tested. The  elastic moduli of the resulting uncoated substrates was measured by flat-punch indentation tests with a cylindrical indenter of diameter $2\rind = 1.25\mathrm{~mm}$, and are plotted in fig.~\ref{fig:SubStiff} as a function of the fraction of crosslinker used for each part of base.

\section*{Appendix B: Details of the solution technique}

\subsection*{Theoretical background}

To solve the pair of integral  equations \eqref{eqn:ptildeND1}--\eqref{eqn:ptildeND2}, we follow Sneddon \cite{Sneddon1965} in setting 
\beq
K\hankel{P}(K)=\Rind^2K\int_0^1\phi(t) \cos K t~\upd t,
\label{eq:sup3}
\eeq to ensure that \eqref{eqn:ptildeND1} is automatically satisfied. (The additional factor of $\Rind^2$ is introduced for later convenience.) Once the function $\phi(t)$ has been computed, the pressure $P(r)$ is immediately given by
\beq
\Rind^{-2} P(r)=\frac{1}{r}\frac{\upd}{\upd r}\left(\int_r^1  \frac{t \phi(t)}{\sqrt{t^2-r^2}} ~\upd t \right)\qquad r\leq1.
\label{eq:sup4}
\eeq
This writing allows us to compute directly the dimensionless indentation force as
\begin{equation}
-F= -2\pi \Rind^{-2} \int_0^1 r P(r)~\upd r =2\pi\int_0^1  \phi(t)~\upd t \;.
\label{eq:sup5}
\end{equation}
The role of the extra factor $\Rind^2$ in Eq.~(\ref{eq:sup3}) is then to simplify the scaling difference between $F$ and $\phi$.

Substituting \eqref{eq:sup3} into \eqref{eqn:ptildeND2} we have that
\beq
\Rind^2\int_0^1 \Xi(R,t)\phi(t)~\upd t=1,\quad R<1,
\label{eq:sup6}
\eeq
where the kernel
\beq
\Xi(R,t)=\int_0^\infty\frac{\cos Kt}{K^3+\Rind^3}J_0(KR)~\upd K.
\label{eq:sup7}
\eeq
The analytical resolution of Eq.~\eqref{eq:sup6} is not possible because the kernel $\Xi(R,t)$ in \eqref{eq:sup7} is not analytically integrable. However, having written the problem in this way facilitates the numerical solution of \eqref{eq:sup6}, as we now demonstrate.

\subsection*{Numerical implementation}
To obtain a numerical solution it is better to transform the integral equation~(\ref{eq:sup6}) to acquire numerical stability. Eq.~(\ref{eq:sup6}) can be rewritten as 
$$\Rind^{2} \frac{\upd}{\upd s}\left\{\int_0^s\frac{R }{\sqrt{s^2-R^2}}\left[\int_0^1 \Xi(R,t) \phi(t)~\upd t\right]~\upd R\right\} =\frac{\upd}{\upd s}\left\{\int_0^s\frac{R}{\sqrt{s^2-R^2}}~\upd R\right\}$$
 Performing the integrals over $R$, one can rewrite this equation as 
\begin{equation}
 \int_0^1 \left\{I[\Rind (s+t)]+I[\Rind|s-t|]\right\} \phi(t)~\upd t=1 \qquad s\leq1
 \label{eq:sup10}
\end{equation}
where 
\begin{equation}
I(x)=\tfrac{1}{2}\int_0^\infty  \frac{ \cos Kx}{K^3+1}~\upd K
 \label{eq:sup11}
\end{equation}
The function $I(x)$ can be written in terms of the Meijer $G$ function and can be evaluated numerically; this shows that its behaviour is  regular for any value of $x$. Therefore the integral equation~\eqref{eq:sup10} with $I(x)$ given by Eq.~\eqref{eq:sup11} may be solved numerically without problems. We discretize the interval $0\leq t\leq1$ to determine a linear system for $\phi$ at various grid points; this linear system is readily solved.


\begin{thebibliography}{10}

\bibitem{Abbott1999}
J.~A. Abbott.
\newblock Quality measurement of fruits and vegetables.
\newblock {\em Postharv. Biol. Tech.}, 15:207--225, 1999.

\bibitem{Abramowitz1964}
Milton Abramowitz and Irene~A. Stegun.
\newblock {\em Handbook of Mathematical Functions with Formulas, Graphs, and
  Mathematical Tables}.
\newblock Dover, New York, 1964.

\bibitem{Boussinesq1878}
J.~Boussinesq.
\newblock \'{E}quilibre d'\'{e}lasticit\'{e} d'un sol isotrope sans pesanteur,
  supportant diff\'{e}rents poids.
\newblock {\em C. R. Acad. Sci.}, 86:1260--1263, 1878.

\bibitem{Chopin2008}
J.~Chopin, D.~Vella, and A.~Boudaoud.
\newblock The liquid blister test.
\newblock {\em Proc. R. Soc. A}, 464(2099):2887--2906, 2008.

\bibitem{Duffy2008}
D.~G. Duffy.
\newblock {\em Mixed Boundary Value Problems}.
\newblock Chapman \& Hall, Boca Raton, 2008.

\bibitem{Duprat1995}
F.~Duprat, M.-G. Grotte, E.~Pietri, and C.J. Studman.
\newblock A multi-purpose firmness tester for fruits and vegetables.
\newblock {\em Comput. Electr. Agri.}, 12:211--223, 1995.

\bibitem{Emadi2005}
B.~Emadi, V.~Kosse, and P.~K. Yarlagadda.
\newblock Elastic contact versus indentation modeling of multi-layered
  materials.
\newblock {\em Int. J. Food Prop.}, 8:277--287, 2005.

\bibitem{Gao1992}
H.~J. Gao, C.~H. Chiu, and J.~Lee.
\newblock Elastic contact versus indentation modeling of multi-layered
  materials.
\newblock {\em Int. J. Solids Struct.}, 29:2471--2492, 1992.

\bibitem{Grotte2001}
M.~Grotte, F.~Duprat, D.~Loonis, and E.~Pi\'{e}tri.
\newblock Mechanical properties of the skin and the flesh of apples,.
\newblock {\em Int. J. Food Prop.}, 4:149--161, 2001.

\bibitem{Harding1945}
J.~W. Harding and I.~N. Sneddon.
\newblock The elastic stresses produced by the indentation of the plane surface
  of a semi-infinite elastic solid by a rigid punch.
\newblock {\em Math. Proc. Camb. Phil. Soc.}, 41:16--26, 1945.

\bibitem{Johnson1987}
K.~L. Johnson.
\newblock {\em Contact Mechanics}.
\newblock Cambridge University Press, Cambridge, 1987.

\bibitem{Johnston2014}
D.~Johnston, D.~K. McCluskey, C.~K.~L. Tan, and M.~C. Tracey.
\newblock Mechanical characterization of bulk {S}ylgard 184 for microfluidics
  and microengineering.
\newblock {\em J. Micromech. Microeng.}, 24:035017, 2014.

\bibitem{Kaltenbrunner2013}
M.~Kaltenbrunner, T.~Sekitani, J.~Reeder, T.~Yokota, K.~Kuribara, T.~Tokuhara,
  M.~Drack, R.~Schw\"{o}diauer, I.~Graz, S.~Bauer-Gogonea, S.~Bauer, and
  T.~Someya.
\newblock An ultra-lightweight design for imperceptible plastic electronics.
\newblock {\em Nature}, 499:458--463, 2013.

\bibitem{Karpitschka2016}
S.~Karpitschka, L.~{van Wijngaarden}, and J.~H. Snoeijer.
\newblock Surface tension regularizes the crack singularity of adhesion.
\newblock {\em Soft Matter}, 12:4463--4471, 2016.

\bibitem{Kim2012}
J.~B. Kim, P.~Kim, N.~C. Pegard, S.~J. Oh, C.~R. Kagan, J.~W. Fleischer, H.~A.
  Stone, and Y.-L. Loo.
\newblock {Wrinkles and deep folds as photonic structures in photovoltaics}.
\newblock {\em Nat. Photonics}, 6(5):327--332, 2012.

\bibitem{Li1997}
J.~Li and T.-W. Chou.
\newblock Elastic field of a thin-film/substrate system under an axisymmetric
  loading.
\newblock {\em Int. J. Solid. Struct.}, 34:4463--4478, 1997.

\bibitem{Liu2019}
Y.~Liu, Y.~Wei, and P.~Chen.
\newblock Characterization of mechanical properties of two-dimensional
  materials mounted on soft substrate.
\newblock {\em Int. J. Mech. Sci.}, 151:214--221, 2019.

\bibitem{Love1939}
A.~E.~H. Love.
\newblock Boussinesq's problem for a rigid cone.
\newblock {\em Quart. J. Math.}, 10:161--175, 1939.

\bibitem{Mizrach1992}
A.~Mizrach, D.~Nahir, and B.~Ronen.
\newblock Mechanical thumb sensor for fruit and vegetable sorting.
\newblock {\em Trans. Am. Soc. Agri. Eng.}, 35:247--250, 1992.

\bibitem{Olver2010}
F.~W.~J. Olver.
\newblock {\em NIST handbook of mathematical functions}.
\newblock Cambridge University Press, 2010.

\bibitem{Paraense2017}
M.~O. Paraense, T.~H. {Rodrigues da Cunha}, A.~S. Ferlauto, and K.~C. {de Souza
  Figueiredo}.
\newblock Monolayer and bilayer graphene on polydimethylsiloxane as a composite
  membrane for gas-barrier applications.
\newblock {\em J. Appl. Polymer Sci.}, 134:45521, 2017.

\bibitem{Perriot2004}
A.~Perriot and E.~Barthel.
\newblock Elastic contact to a coated half-space: Effective elastic modulus and
  real penetration.
\newblock {\em J. Mater. Res.}, 19:600--609, 2004.

\bibitem{Plocharski2003}
W.~J. P\l{}ocharski and D.~Konopacka.
\newblock Non-destructive, mechanical method for measurement of plums'
  firmness.
\newblock {\em Int. Agrophysics}, 17:199--206, 2003.

\bibitem{Polderdijk2000}
J.~J. Polderdijk, R.~M. Kho, and A.~P.~M. {de Kruif}.
\newblock Firmness of mangoes measured acoustically mechanically and manually.
\newblock In {\em Proceedings of the Sixth International Mango Symposium},
  pages 861--865. Int. Soc. Hort. Sci., 2000.

\bibitem{Scimeca2019}
L.~Scimeca, P.~Maiolino, D.~Cardin-Catalan, A.~P. {del Pobil}, A.~Morales, and
  F.~Iida.
\newblock Non-destructive robotic assessment of mango ripeness via multi-point
  soft haptics.
\newblock In {\em Proceedings of the 2019 International Conference on Robotics
  and Automation (ICRA)}, pages 1821--1826. IEEE, 2019.

\bibitem{Sneddon1965}
I.~N. Sneddon.
\newblock The relation between load and penetration in the axisymmetric
  boussinesq problem for a punch of arbitrary profile.
\newblock {\em Int. J. Engng Sci.}, 3:47--57, 1965.

\bibitem{Sneddon1995}
I.~N. Sneddon.
\newblock {\em Fourier Transforms}.
\newblock Dover, New York, 1995.

\bibitem{Taffetani2017}
M.~Taffetani and D.~Vella.
\newblock Regimes of wrinkling in pressurized elastic shells.
\newblock {\em Phil. Trans. R. Soc. A}, 375:20160330, 2017.

\bibitem{Ventsel2001}
E.~Ventsel and T.~Krauthammer.
\newblock {\em Thin Plates and Shells}.
\newblock Marcel Dekker, New York, NY, 2001.

\bibitem{Wang2017}
J.~Wang, Q.~Cui, H.~Li, and Y.~Liu.
\newblock Mechanical properties and microstructure of apple peels during
  storage.
\newblock {\em Int. J. Food. Prop.}, 20:1159--1173, 2017.

\bibitem{Wang2014}
Z.~Wang, A.~A. Volinsky, and N.~D. Gallant.
\newblock Crosslinking effect on polydimethylsiloxane elastic modulus measured
  by custom-built compression instrument.
\newblock {\em J. Appl. Polym. Sci.}, 131:41050, 2014.

\bibitem{Yu1990}
H.~Y. Yu, S.~C. Sanday, and B.~B. Rath.
\newblock The effect of substrate on the elastic properties of films determined
  by the indentation test --- axisymmetric {B}oussinesq problem.
\newblock {\em J. Mech. Phys. Solids}, 38:745--764, 1990.

\end{thebibliography}

\end{document}